\documentclass[12pt]{iopart}
\usepackage{cite}
\usepackage{graphicx} 
\usepackage{booktabs}
\begin{document}

\title{Nucleon charge radius measurement with low-energy electron scattering}

\author{Clement Legris$^1$, Rika Danjo$^1$, Taiga Goke$^1$, Yuki Honda$^1$, Kengo Hotta$^1$, Rin Kagami$^2$, Hiroki Kobayashi$^2$, Michael Kohl$^3$, Yukie Maeda$^4$, Dominique Marchand$^5$, Edward Morris$^6$, Toshiya Muto$^1$, Tetsuya Ohnishi$^7$, Shunto Sasaki$^1$, Toshimi Suda$^1$, Kyo Tsukada$^2$, Eric Voutier$^5$, Toyohiro Yamauchi$^1$ and Kosei Yoshimoto$^1$}

\address{$^1$Research center for Advanced Radioisotope Science (RARiS), Tohoku University, Sendai, Japan}
\address{$^2$Institute for Chemical Research (ICR), Kyoto University, Japan}
\address{$^3$Hampton University, Department of Physics, USA}
\address{$^4$Miyazaki University, Faculty of Engineering, Japan}
\address{$^5$Paris-Saclay University, CNRS/IN2P3, IJCLab, 91405 Orsay, France}
\address{$^6$University of Exeter, United Kingdom}
\address{$^7$RIKEN Nishina Center for Accelerator-Based Science, Japan}
\ead{legris@raris.tohoku.ac.jp}
\vspace{10pt}
\begin{indented}
\item[] 26 July 2025
\end{indented}

\begin{abstract}
The charge radius is one of the most basic characteristics of the nucleons. The proton charge radius is especially of great importance for many applications such as the structure studies of the atomic nuclei, the determination of the Rydberg constant and QED tests. Its determination is thus a hot topic in several physics communities due to inconsistent results using electron scattering, atomic and muonic hydrogen spectroscopy. A new measurement of the proton and deuteron charge radii with low energy electron scattering is being conducted in the Research Center for Accelerator and Radioisotope Science (RARiS), Tohoku University, Japan. The current status of the experiment is discussed in the present paper.
\end{abstract}

\begin{indented}
\item[]
\end{indented}

\section{Introduction}

Proton and neutron are the building blocks of visible baryonic matter. While some characteristics such as their masses are determined with a great precision, others are still active research subjects, for example proton decay or neutron lifetime \cite{protondecay, neutronlifetime}. Among them, the proton radius is currently a hot topic in the nuclear and atomic physics communities. 

\begin{figure}[ht]
    \centering
    \includegraphics[width=0.9\linewidth]{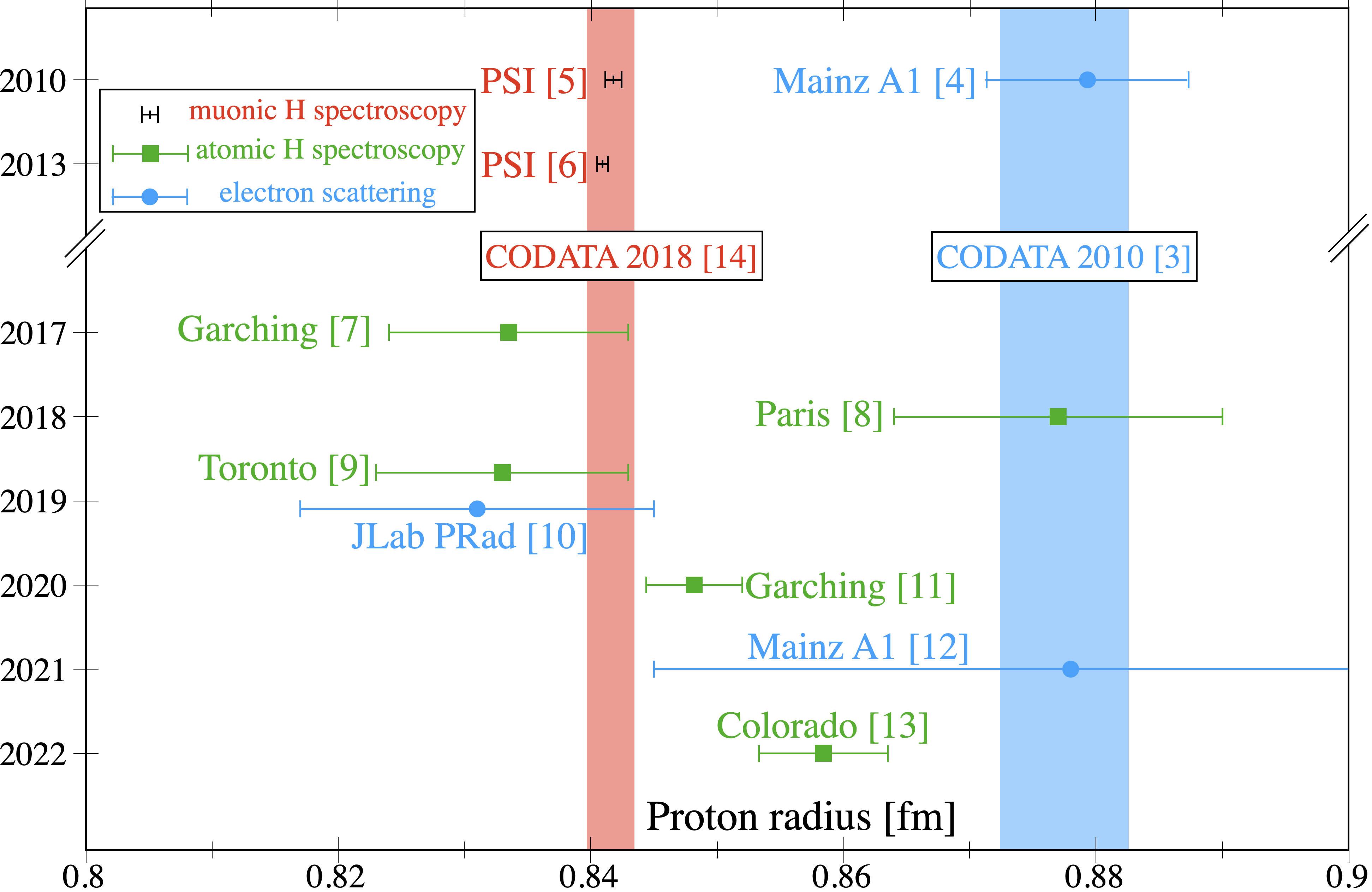}
    \caption{Current experimental status of the proton charge radius puzzle.}
    \label{prad_puzzle}
\end{figure}

Measurements of the proton charge radius were conducted before 2010 using electron scattering as well as atomic hydrogen spectroscopy and reported that the proton charge radius was consistent with 0.8775(51) fm, the value recommended by CODATA 2010\cite{codata2010,mainz2014}. However, in 2010 a high-accuracy experiment using muonic hydrogen spectroscopy measured a radius of 0.84184(67) fm, in disagreement by about 7$\sigma$ with the CODATA recommended value \cite{pohl2010}. In 2013, the muonic hydrogen spectroscopy result was confirmed by the same group using the same experimental setup\cite{antognini2013}. Since then, many experiments were carried out using electron scattering and atomic hydrogen spectroscopy. Most of the published results, as well as data reanalysis and theoretical studies, are either close to 0.84 or 0.88 fm, as shown in figure \ref{prad_puzzle} \cite{beyer2017, fleurbaey2018,bezginov2019,prad2019,grinin2020,miha2021,brandt2022,codata2018}. 

Inconsistencies between several results were pointed out: results of hydrogen spectroscopy, \cite{fleurbaey2018} and \cite{grinin2020}, point to different radii measuring the same transition (1S-3S), which may suggest unclarified systematic uncertainties, and the two most recent electron scattering experiments reported different radii as well as some inconsistencies \cite{mainz2014, prad2019, atoms, 9years, NRP}. Many possible explanations for these discrepancies were investigated: missing terms in the theoretical description of muonic hydrogen spectroscopy, inconsistent definitions of the proton charge radius between electron scattering and spectroscopy, lepton non-universality, etc \cite{carlson2015}. Considerable theoretical work led to the conclusion that there is no missing term in the muonic hydrogen spectroscopy theory accounting for the discrepancy \cite{carlson2015}. The definition of the proton charge radius is also consistent between electron scattering, atomic and muonic hydrogen spectroscopy \cite{miller2019}.  The accuracy and the reliability of several experiments are therefore questioned and more data are necessary to understand these inconsistencies and to solve this issue, especially at low momentum transfer \cite{atoms, 9years, NRP, rn}. The ULQ$^2$ (Ultra-Low momentum transfer $Q^2$) experiment is currently carrying out the most accurate and reliable determination of the proton charge radius using low-energy electron scattering and very low $Q^2$ with a targeted accuracy  of 1 \% for the proton charge radius.

\section{Determination of the proton charge radius with electron scattering}

Electron scattering is one of the best tools to access nuclear charge distribution or, equivalently, nuclear form factors \cite{suda2017}. Electron is a point-like particle that interacts through the electromagnetic interaction, which can be calculated exactly to any order with QED. As a consequence, the nucleus-electron scattering cross-section can be calculated with a very high precision.

\subsection{General description} 

Under the one-photon-exchange (OPE) approximation and neglecting the electron mass, the elastic electron-proton cross-section is,

\begin{equation}
	\left(\frac{d \sigma}{d \Omega}\right) = 
	\left(\frac{d \sigma}{d \Omega}\right)_{Mott} f_{recoil}\frac{G_{E}^2(Q^2)+\frac{\tau}{\epsilon}G_{M}^2(Q^2)}{1+\tau}.
        \label{cross-section1}
\end{equation}	

where $\left(\frac{d \sigma}{d \Omega}\right)_{Mott}$ corresponds to the Mott cross-section, the cross-section for a point-like particle, $f_{recoil}$ is the ratio of the energy of the electron after proton recoil and its energy before scattering, $\tau = \frac{Q^2}{4M_p^2}$ and $\epsilon = [1+ 2(1+\tau)\tan^2{\theta/2}]^{-1}$ is the virtual photon polarization. The form factors, $G_E$ and $G_M$, denote, respectively, the Sachs electric and magnetic form factors and describe the charge and magnetic distribution of the proton as a function of $Q^2 = -q^2$, the opposite of the 4-momentum transfer squared. In the ULQ$^2$ experiment, the electron energy is very low thus the electron mass cannot be neglected. An expression of the cross-section including the electron mass can be found in \cite{scattmuon}. 

The proton charge radius $r_p$ is defined by the slope of the electric form factor in the limit of $Q^2 \rightarrow 0$ \cite{miller2019}, namely,

\begin{equation}
	r_p = \sqrt{-6\hbar^2 \lim_{Q^2 \rightarrow 0} \frac{dG_E}{dQ^2}}.
\end{equation}

Measurement at $Q^2 = 0$ is impossible thus extracting the proton radius requires taking data at small $Q^2$.

\subsection{ULQ$^2$ key concepts}

\begin{figure}[ht]
    \centering
    \includegraphics[width=0.9\linewidth]{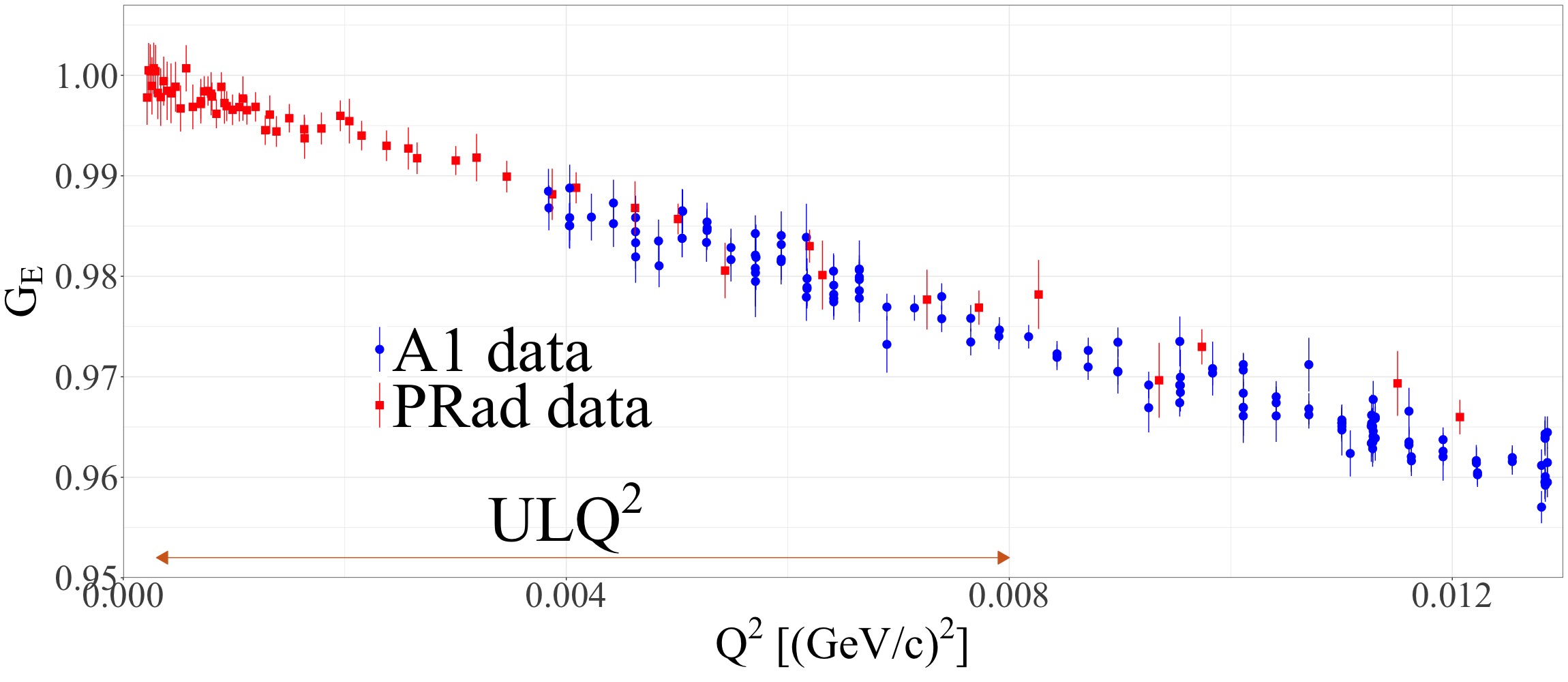}
    \caption{Measured proton electric form factor $G_E$ using A1 and PRad data within the ULQ$^2$ kinematical range.}
    \label{GE}
   \end{figure}

The ULQ$^2$ experiment is carried out in RARiS, Tohoku University, using the 60-MeV-accelerator, which delivers any energy between 10 and 60 MeV. The measurement can be carried at any scattering angle between 30$^\circ$ and 150$^\circ$. Therefore, the accessible $Q^2$ region by the ULQ$^2$ experiment is $3 \times 10^{-5}$ - $1.2 \times 10^{-2}$ (GeV/$c$)$^2$. During the experiment, data taking was conducted within the $3 \times 10^{-4}$ - $8 \times 10^{-3}$ (GeV/$c$)$^2$ region, which corresponds to the lowest-ever reached $Q^2$ except for the PRad experiment at the Jefferson Laboratory \cite{prad2019}. As shown in figure \ref{GE}, within the data taking $Q^2$ range of the ULQ$^2$ experiment, the proton electric form factor $G_E$ ranges from 0.97 to 1.  To determine the proton charge radius with 1\% accuracy from this low $Q^2$ region, the required accuracy of the cross-section is thus 0.1 \%.

In order to reach the required accuracy of 0.1 \%  for the elastic electron-proton scattering, absolute cross-section measurement is performed using a polyethylene (CH$_2$) target during the ULQ$^2$ experiment. Electron-proton and electron-carbon scattering events are detected simultaneously throughout the experiment. The momentum of the scattered electrons depends on the mass $M_X$ of the nucleus involved in the interaction with the following relation (neglecting the electron mass),

\begin{equation}
	p^\prime_\mathrm{X} = \frac{p}{1+\frac{2p\sin^2{\theta/2}}{M_\mathrm{X}}}.
	\label{pscatt}
\end{equation}
 
 The $^{12}$C nucleus is about 12 times heavier than proton thus the momenta of scattered electrons on proton and carbon are different and in the data taking $Q^2$ region, their relative difference is within a few \%. Therefore, thanks to the high-resolution equipment and large momentum-bite, the electron-carbon and electron-hydrogen elastic scattering peaks are simultaneously detected within the spectrometer momentum acceptance and well-separated, see figure \ref{2d}. The electron-proton cross-section is expressed with regards to the electron-carbon cross-section,

\begin{equation}
	\left( \frac{d\sigma}{d\Omega}\right)_{e-p} = \frac{N_{e-\mathrm{p}}/N_{e-^{12}\mathrm{C}}}{n_\mathrm{p}/n_{^{12}\mathrm{C}}}
	\left( \frac{d\sigma}{d\Omega}\right)_{e-^{12}\mathrm{C}}.
\end{equation}

$N_{e-\mathrm{X}}$ denotes the number of detected e-X scattering events and $n_\mathrm{X}$ the number of X nuclei in the target. This expression cancels out the uncertainty related to the beam dose and the detector acceptance. Elastic $e-^{12}$C scattering has been extensively studied \cite{cardman1980,reuter1982,offermann1991} and the $^{12}$C charge radius has been measured with an accuracy better than 0.1 \% using muonic $^{12}$C \cite{ruckstuhl1984}. As a consequence, $^{12}$C form factor as well as proton electric form factor can be determined with 0.1 \% accuracy in the ULQ$^2$ $Q^2$ range. 

 The experiment was designed to perform Rosenbluth separation if necessary: several measurements of the electron-proton cross-section are carried out with a fixed $Q^2$ but with different scattering angles $\theta$ to extract separately the electric and magnetic form factors. However, at such low $Q^2$ the magnetic term contribution $\frac{\tau}{\epsilon}G_M^2(Q^2)$ is very small. Since the proton magnetic form factor can be well approximated by the dipole form factor, $\frac{G_M(Q^2)}{\mu_p} \sim (1+\frac{Q^2}{0.71 (\mathrm{GeV}/c)^2})^{-2}$ with $\mu_p \approx 2.793$ the proton magnetic moment \cite{GM}, the magnetic term contribution goes from  $6.6 \times 10^{-4}$ at $3 \times 10^{-4}$ (GeV/$c$)$^2$ to $1.7 \times 10^{-2}$ at $8 \times 10^{-3}$ (GeV/$c$)$^2$. Thus, a simple model of the magnetic form factor suffices to correctly estimate the magnetic term and a Rosenbluth separation is redundant.

\section{The ULQ$^2$ experimental setup}

\subsection{ULQ$^2$ beam line and magnetic twin spectrometers}

The ULQ$^2$ experiment is using the 60-MeV linear electron accelerator at Tohoku University's RARiS facility, which has been primarily used for radioisotope production. It underwent many improvements to fulfill the requirements of the ULQ$^2$ experiment. It can currently provide a pulsed beam with a repetition of 300Hz, a beam duty of 0.1 \% and an intensity up to 400 nA. At the target position, the beam position spread is about 1 mm and the beam momentum spread 0.1 \%.

\begin{figure}[ht]
    \centering
    \includegraphics[width=0.7\linewidth]{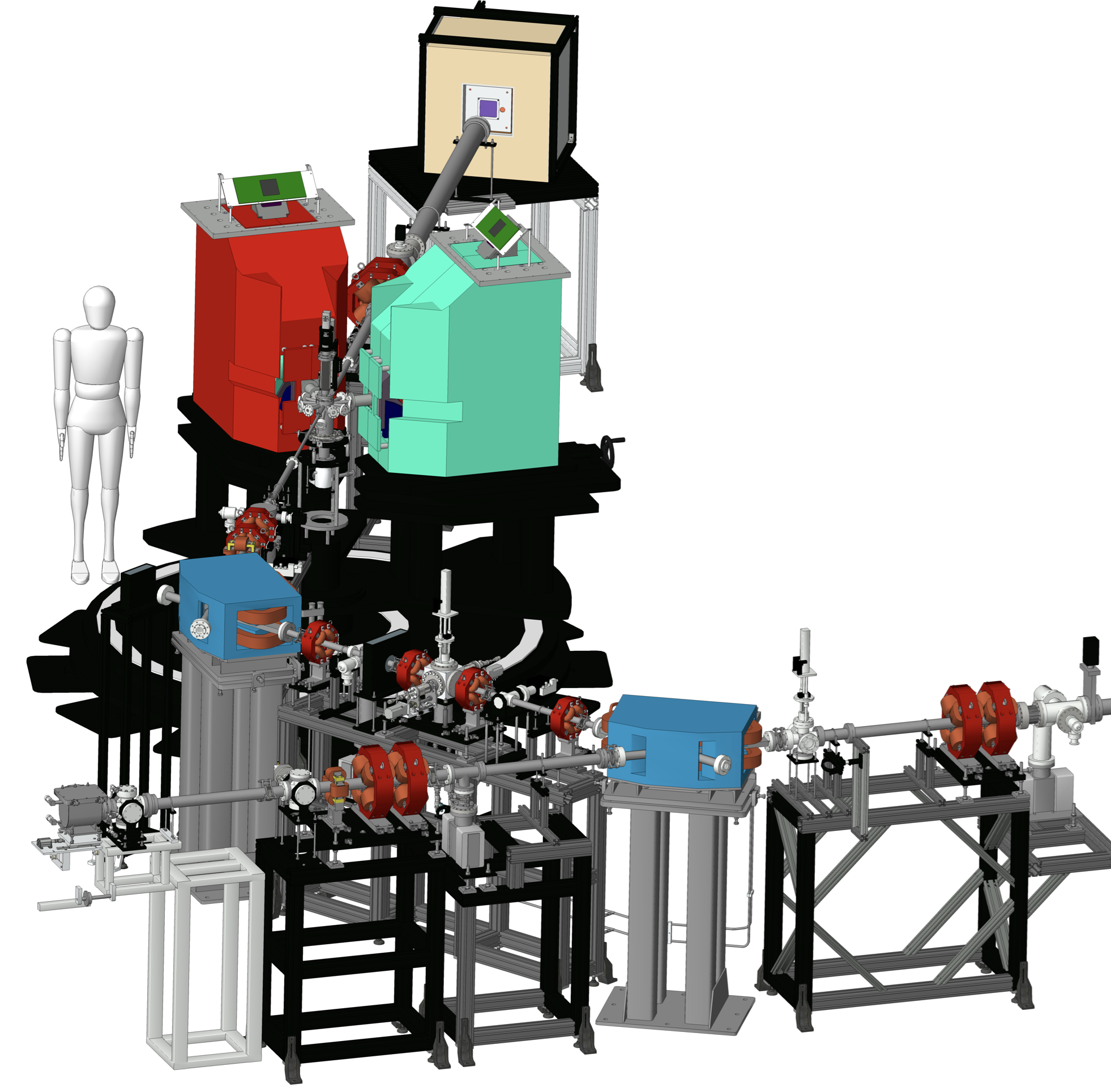}
    \caption{Schematic layout of the ULQ$^2$ beam line and the twin spectrometers.}
    \label{beamline}
\end{figure}

\begin{figure}[ht]
    \centering
    \includegraphics[width=0.5\linewidth]{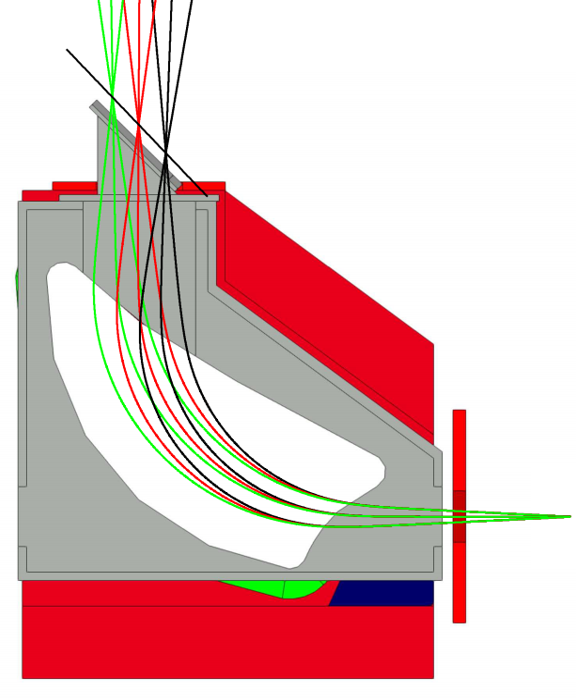}
    \caption{Scattered electrons focused in the focal plane of the spectrometer.}
    \label{spectrometer}
\end{figure}

The ULQ$^2$ experiment uses two high-resolution magnetic spectrometers built with identical design. The first spectrometer was installed in 2019 and has been partially commissioned in November 2020 while the second one was installed in August 2021, see figure \ref{beamline}. The twin spectrometers are dipole magnets used to measure the momentum and the scattering angle of the electrons scattered off by the target. Due to the extremely-low momentum of the electrons, multiple scattering makes tracking with several detectors at the focal plane impossible. Instead, the spectrometers were carefully designed to focus scattered electrons along the momentum direction, as shown in figure \ref{spectrometer}, and defocus along the transverse direction, to increase angular resolution, in their focal plane, where detectors are located. The momentum and the scattering angle of the electrons can be directly obtained from their 2D position on the detectors.

The twin spectrometers can sustain a current up to 300 A, creating a magnetic field up to 0.4 T and have distinct usages. Hydrogen nuclei in the target gradually escape from the target due to beam irradiation leading to a progressive evolution of the hydrogen-carbon ratio in the target.  The first spectrometer measures the electron-proton and electron-carbon scattering yield ratio. The second spectrometer acts as a target status monitor and records the electron-proton and electron-carbon yields to correct the evolution of the hydrogen-carbon yield ratio. 

\subsection{Detectors}

\begin{figure}[ht]
    \centering
    \includegraphics[width=0.8\linewidth]{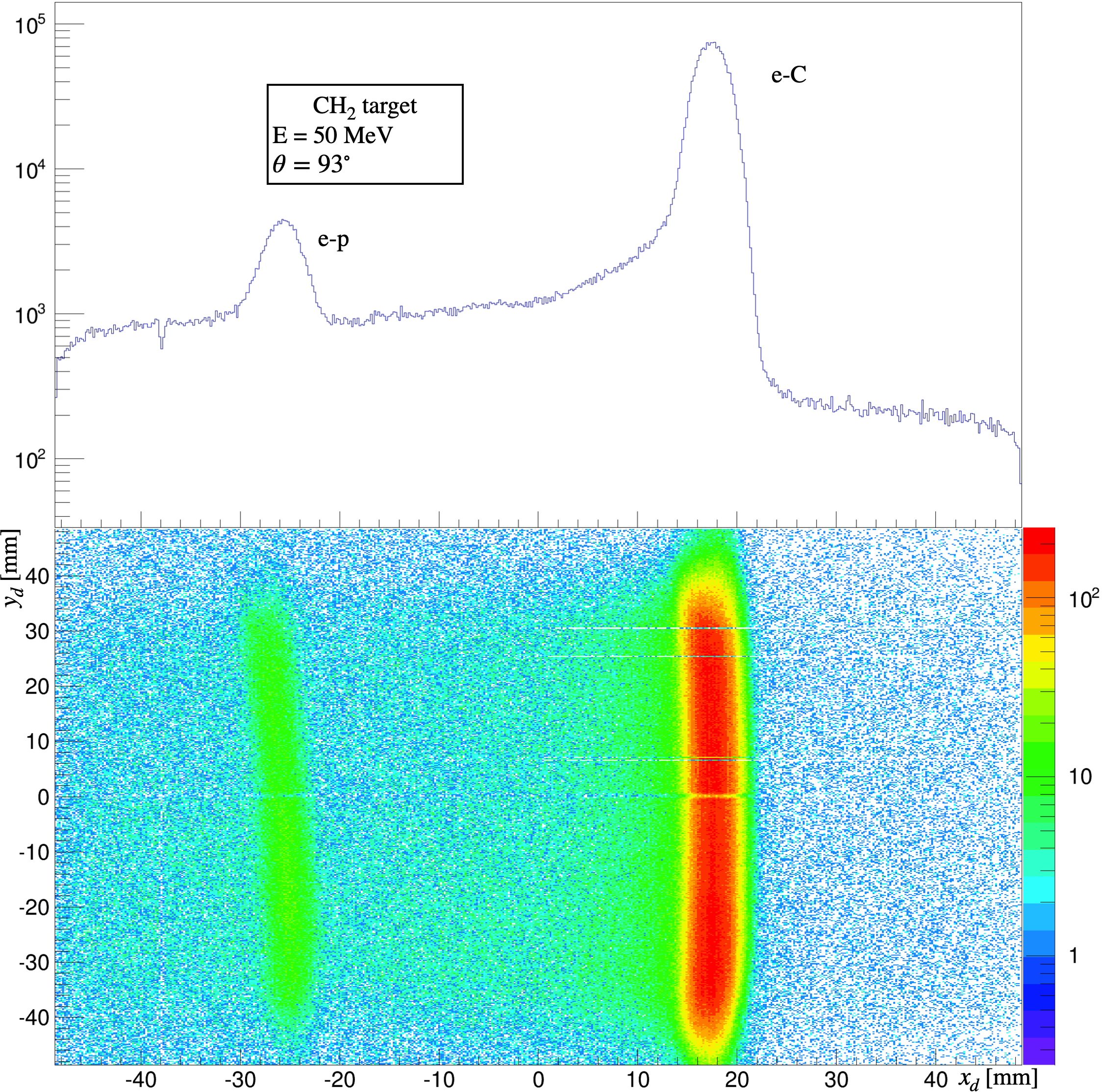}
    \caption{Typical $(x_d,y_d)$ 2D spectrum obtained with the pair of SSSDs (bottom) and the projection along $x_d$ (up).}
    \label{2d}
\end{figure}

The detectors used for the ULQ$^2$ experiment are fully digitized onboard Single-sided Silicon Strip Detectors (SSSDs)  \cite{aoyagiSSD} developed with the J-PARC muon g-2/EDM collaboration \cite{g-2}. The surface of each SSSD is composed of a 100$\times$100 mm$^2$ silicon sensor between 2 boards. Each board contains Application Specific Integrated Circuits (ASIC), connected to 512 channels. Each channel is 0.19-mm wide and 0.32-mm thick.  Two SSSDs are placed in the focal plane of each spectrometer: one SSSD measures  $x_d$ while the other measures $y_d$. A typical e-CH$_2$ experimental spectrum is displayed in figure \ref{2d}. The electron-carbon and electron-hydrogen scattering peaks are well separated thanks to the very high-resolution of the spectrometers.\\

The 2D $(x_d,y_d)$ position on the detectors is related to the beam position on the target, the in-plane scattering angle $\theta$ and the momentum of the scattered electron $p^\prime$. Excluding the beam position-related and small-effect terms, 

\begin{equation}
	x_d = (x_d|\delta) \delta,
	\label{momdisp}
\end{equation}

\begin{equation}
	y_d = [(y_d|\Delta \theta)  + (y_d|\Delta \theta \delta) \delta]\Delta \theta .
	\label{thetadisp}
\end{equation}

where $\Delta \theta = \theta - \theta_{SP}$ denotes the difference between the scattering angle and the spectrometer angle and $\delta$ represents the variation of momentum from $x_d = 0$.  The latter can be expressed as a function of the momentum of the scattered electron $p^\prime$ or the magnetic field of the spectrometer $B_{SP}$: $\delta = \frac{p^\prime -p_{SP}}{p_{SP}} = \frac{B^\prime - B_{SP}}{B_{SP}}$. The precise determination of the momentum dispersion is crucial for the beam energy determination. In order to obtain $Q^2$ with an accuracy better than $10^{-3}$, the beam momentum is calculated from the e-CH$_2$ spectrum. Using equations (\ref{pscatt}) and (\ref{momdisp}),

\begin{equation}
	\frac{p^\prime_\mathrm{C}}{p^\prime_\mathrm{p}} = \frac{1+\frac{2p \sin^2 {\theta/2}}{M_\mathrm{p}}}{1+\frac{2p\sin^2{\theta/2}}{M_\mathrm{C}}} = \frac{1+\frac{x_\mathrm{C}}{(x_d|\delta)}}{1+\frac{x_\mathrm{p}}{(x_d|\delta)}}.
\end{equation}

The $x_d$-positions of the elastic electron-carbon and electron-proton scattering peaks are $x_\mathrm{C}$ and $x_\mathrm{p}$ respectively so at first order, the beam momentum is proportional to the distance between the peaks $x_\mathrm{C} -x_\mathrm{p}$. Therefore, the matrix elements were carefully determined during commissioning. In particular,  the momentum dispersion $(x_d|\delta) \approx 865$ mm was measured with an accuracy better than 0.1 \% at the beginning of each data taking period by varying $B_{SP}$ while keeping the beam energy constant, monitored by the other spectrometer. The angular dispersions $(y_d|\Delta \theta) \approx 1$ mm/mrad and $(y_d|\Delta \theta \delta) \approx 2$ mm/mrad were measured using a sieve slit at the entrance of the spectrometers. The momentum and in-plane angular detection acceptances are about 11 \% and  6$^\circ$ respectively. 

The hydrogen and carbon scattering peaks have to be well-separated at any momentum transfer, especially at the lowest momentum transfer: $Q^2 = 3\times 10^{-4}$ (GeV/$c$)$^2$. For that purpose, the spectrometer momentum resolution $\frac{\Delta p}{p}$ have to be better than 0.1 \%. This requirement is fulfilled by both spectrometers as it was confirmed during the commissioning phase carried out in 2021. The measured spectrometer resolution $\frac{\Delta p}{p}$ is about $6 \times 10^{-4}$ and the angular resolution is close to 1 mrad. 

The beam position-related and small-effect terms not shown in equations (\ref{momdisp}) and (\ref{thetadisp}) were also determined during the commissioning phase. A shift of the beam position vertically and horizontally on the target results in a shift of the same order of magnitude along $x_d$ and $y_d$, respectively. The small-effect terms are negligible close to the center of the SSSDs but modify $x_d$ by a few hundreds of $\mu$m on the edges of the SSSDs.

\section{Deuteron and neutron radii determination}

\subsection{Deuteron charge radius puzzle}

\begin{figure}[ht]
    \centering
    \includegraphics[width=0.9\linewidth]{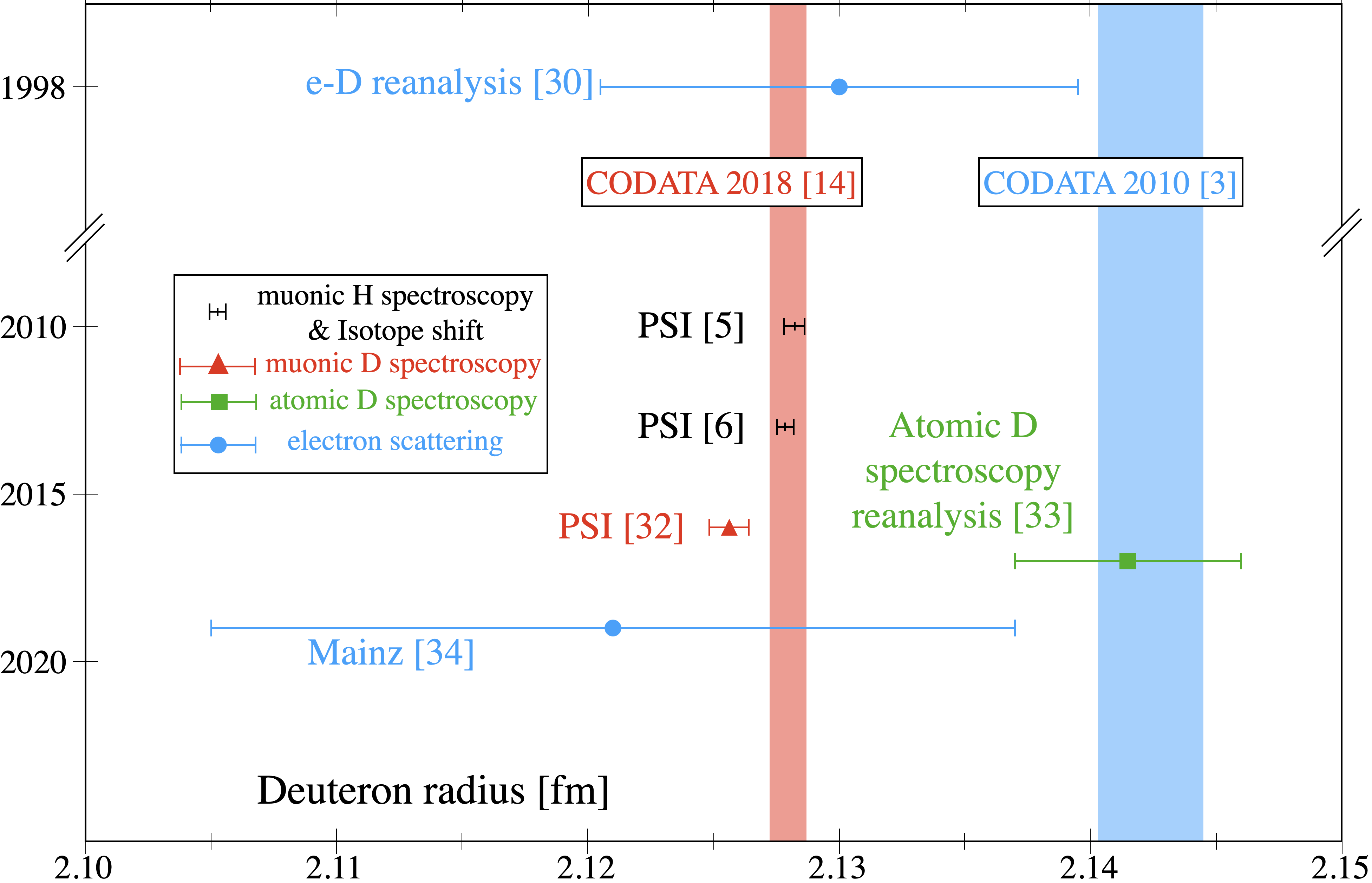}
    \caption{Current experimental status of the deuteron charge radius puzzle.}
    \label{drad_puzzle}
\end{figure}

In parallel to the proton charge puzzle, the deuteron charge radius is also facing inconsistent results between muonic and atomic deuterium spectroscopy. An analysis of the electron-deuteron data prior to 1998 reported a radius of 2.130(9) fm \cite{sick1998}, in slight disagreement with the atomic deuterium spectroscopy radius of 2.1424(21) fm recommended by CODATA 2010 \cite{codata2010}. Proton and deuteron radii are very strongly correlated to each other: the isotope shift $\langle r_d^2 \rangle - \langle r_p^2 \rangle$ is known with a very high accuracy \cite{isotope_shift}. Thus the small proton radius measured with muonic hydrogen spectroscopy implies a smaller deuteron radius, close to 2.128 fm and inconsistent with atomic deuterium spectroscopy,  as shown in figure \ref{drad_puzzle}. A recent measurement of the deuteron radius using muonic deuterium spectroscopy reported a smaller radius \cite{pohluD2016} whereas a reanalysis of the atomic deuterium spectroscopy data led to a larger radius \cite{pohleD2017}. The deuteron form factors were also measured at Mainz University with electron scattering \cite{mainz_ed}. The result slightly favors the small radius.

Not only the deuteron radius value is not precisely known but electron-deuteron scattering data are missing as compared to electron-proton scattering. Except for the recent measurement of the deuteron form factors performed at MAMI (Mainz Microtron), the electron-deuteron scattering experiments were conducted between 1973 and 1990 and the data are normalized to the electron-proton cross-section \cite{berard1973,simon1981,platchkov1990}. Moreover, only relative cross-section were carried out and the data were taken at relatively high $Q^2$: data with $Q^2$ smaller than 0.01 (GeV/$c$)$^2$ are very scarce. 

To provide reliable and high-accuracy data at extremely low $Q^2$, the ULQ$^2$ experiment is conducting the measurement of the electron-deuteron cross-section with a momentum transfer squared between $7.5 \times 10^{-4}$ and $7.5 \times 10^{-3}$ (GeV/$c$)$^2$ using a deuterated polyethylene CD$_2$ target. Similarly to the electron-proton ULQ$^2$ experiment, the high-resolution spectrometers can distinguish the electron-deuteron and electron-proton scattering peaks to determine the absolute electron-deuteron cross-section to conduct with electron-deuteron scattering the same procedure as with electron-proton scattering.

\subsection{Neutron charge radius}

The mean-square neutron charge radius $\langle r_n^2 \rangle= -0.1155(17)$ fm$^2$ recommended by the Particle Data Group (PDG) is derived from measurements of the neutron-electron scattering length  \cite{pdg2024}. Several experiments were conducted between 1973 and 1997 by irradiating a liquid Pb or Bi target with a neutron beam. The mean-square neutron charge radius was derived from the scattering length, obtained from the total cross section measured by neutron transmission. In 2020, a precise theoretical calculation of the deuteron structure radius with chiral effective field theory ($\chi$EFT) led to a larger mean-square neutron charge radius $\langle r_n^2  \rangle = -0.106(9)$ fm$^2$ using the following relation \cite{xEFT}:

\begin{equation}
	\langle r_d^2 \rangle= r_{str}^2  + \langle r_p^2 \rangle + \langle r_n^2 \rangle +\frac{3}{4m_p^2}.
\end{equation}

The mean-squared structure radius $r_{str}^2$ corresponds to the deuteron radius assuming that nucleons are point-like particles.  Moreover, in 2021, an interferometry study reported a larger value $\langle r_n^2 \rangle = -0.1101(89)$ fm$^2$ also pointing towards a larger mean-square neutron radius. 

As the current ULQ$^2$ experiment will independently measure the proton and deuteron charge radii, it may provide a unique and model-independent value on the mean-squared neutron charge radius. In addition to the deuteron charge radius, electron-deuteron data will also give access to the fourth moment of the charge density, which is also related to the mean-square neutron radius \cite{rn,r4}: 

\begin{equation}
	\langle r_d^4 \rangle = 60\hbar^4 \lim_{Q^2\rightarrow 0} \frac{d^2G_E(Q^2)}{d^2(Q^2)}, 
\end{equation}

\begin{equation}
	\langle r_d^4 \rangle = r_{str}^4 + \frac{10}{3}r_{str}^2 (\langle r_p^2 \rangle + \langle r_n^2 \rangle ) + \textrm{rel. corr.}\ .
\end{equation}

The measurement of the electron-deuteron scattering cross-section at low $Q^2$ will add additional constraints on the mean-square neutron charge radius or on the sum of the proton and neutron mean-square charge radii $\langle r_p^2 \rangle + \langle r_n^2 \rangle $.

\section{Current status and summary}

The measurement of the electron-proton and electron-deuteron cross-sections have already been completed during 2023 and 2024 in RARiS. In total, 18 $Q^2$ data points between $3 \times 10^{-4}$ and $8 \times 10^{-3}$ (GeV/$c$)$^2$ for electron-proton scattering and 17 $Q^2$ data points between $7.5 \times 10^{-4}$ and $7.5 \times 10^{-3}$ (GeV/$c$)$^2$ for electron-deuteron scattering have been taken. Analysis is currently ongoing, results will be released soon.

In this paper, the current status of the proton and deuteron charge radius puzzles was described and a new method of deriving the neutron charge radius was addressed. In this context, the ULQ$^2$ experiment will provide new,  independent and reliable measurements of the proton and deuteron charge radii with electron-scattering in the extremely low momentum transfer region, using the worldwide lowest ever beam-energy for electron scattering. 

\ack
This work is supported by JSPS KAKENHI Grants No. 16H06340 and 20H05635 as well as the Graduate Program on Physics for the Universe (GP-PU), Tohoku University. It also benefited from funding from the European Union’s Horizon 2020 research and innovation program under grant agreement No 824093.

\section*{References}
\bibliography{Bibliography}
\bibliographystyle{iopart-num}

\end{document}